\documentclass[aps,prc,twocolumn,superscriptaddress,nofootinbib]{revtex4-1}
\usepackage{latexsym}
\usepackage{amsmath}
\usepackage{amssymb}
\usepackage{amsfonts}
\usepackage{bm}
\usepackage{physics}
\usepackage{booktabs}

\usepackage{color}
\definecolor{purple}{rgb}{0.5,0,0.5}
\definecolor{blue}{rgb}{0.0,0,0.9}
\definecolor{prdblue}{rgb}{0.133,0.118,0.498}
\usepackage[colorlinks=true, pdfstartview=FitV, linkcolor=prdblue, citecolor= prdblue, urlcolor=prdblue]{hyperref}

\usepackage{supertabular}
\usepackage{placeins}
\usepackage{epsfig}
\usepackage{graphicx}

\usepackage{soul}

\begin{document}
\title{Revisiting p-$^{11}$B Fusion: Updated Cross-sections, Reactivity, and Energy Balance}

\author{Hong-Yi Wang}
\affiliation{School of Physics, Nanjing University, Nanjing, Jiangsu 210093, China}

\author{Yu-Qi Li}
\affiliation{School of Physics, Nanjing University, Nanjing, Jiangsu 210093, China}

\author{Qian Wu}
\affiliation{School of Physics, Nanjing University, Nanjing, Jiangsu 210093, China}

\author{Zhu-Fang Cui}
\email[Contact author: ]{phycui@nju.edu.cn}
\affiliation{School of Physics, Nanjing University, Nanjing, Jiangsu 210093, China}

\date{\today}

\begin{abstract}
Recent experimental progress has substantially improved the available cross-section data for the p-$^{11}$B fusion reaction, particularly in energy regions that previously lacked direct measurements. In this study, we develop a high-precision analytical parameterization of the p-$^{11}$B reaction cross-section over the 0--10 MeV energy range, incorporating the new experimental data into a continuous and numerically efficient representation. Using this parameterization, we evaluate the thermonuclear reactivity of the p-$^{11}$B reaction and examine the effects of the dominant resonance at 0.6 MeV and a newly observed resonance around 4.7 MeV. Furthermore, we assess the energy balance by analyzing the fusion power density and the electron bremsstrahlung power density. Our results indicate that p-$^{11}$B fusion is not precluded by bremsstrahlung constraints when contemporary cross-section data and self-consistent thermal modeling are employed.
\end{abstract}

\maketitle



\section{Introduction}
Today, most energy consumed worldwide is extracted from fossil fuels, which unfortunately not only have negative consequences for the environment but are also finite. In this case, although there are still many challenges to this day, both scientifically and engineering-related, the quest for controlled nuclear fusion energy has been a long-standing dream for humanity since the 1950s, mainly because of its carbon-free nature, high unit energy, and abundance of resources~\cite{IAEA2025}.

Among them, the proton-boron (p-$^{11}$B) fusion, which has been studied since the 1930s and thought as the holy grail~\cite{Oliphant_Rutherford_1933,Dearnaley1957,Segel1965,Kamke1967,moreau_1976,Volosov_2006,Last2011,Labaune:2013dla,Geser2020}, is a very attractive aneutronic fusion reaction pathway, and has received renewed attention as well as made substantial progress in recent years, for example, Refs.~\cite{Margarone2022,Liu2023,Magee2023,Lindsey2024,Liu2024,Ochs2024,Ogawa_2024,Sciscio2025,Liu2025}. Its dominant reaction
\[
\mathrm{p} + {}^{11}\mathrm{B} \rightarrow 3\alpha + 8.7~\mathrm{MeV}
\]
produces only charged $\alpha$ ($^{4}$He) particles and no prompt high-energy neutrons, offering the prospect of reduced material activation and the possibility of direct energy conversion~\cite{Beckman1953,Moreau1977}; 
while side reactions, such as 
\[
\mathrm{p} + {}^{11}\mathrm{B} \rightarrow n+ {}^{11}\mathrm{C}
\]
that produce more manageable low-energy neutrons ($\sim$ 2.765 MeV), only has a fraction around $10^{-5}$, so that makes it an environmentally friendly and cost-effective manner. In addition, the reactants are abundant, non-toxic, non-radioactive, and stable in nature, avoiding the radiological handling challenges associated with conventional deuterium-tritium (D-T) or other fusion fuels \cite{Dawson1981}. Last but not least, it is possible that its equilibrium thermal fusion rate at high temperatures is comparable to that of D-T~\cite{Nevins2000}.

Despite these appealing features, however, the practical feasibility of p-$^{11}$B fusion has long been questioned. The main reason is, owing to its relatively low reactivity, ignition (where the plasma is sustained by the fusion reactions alone) requires ion temperatures on the order of several hundred keV, significantly higher than those for D-T fusion ($\sim$ 20 keV)~\cite{Hartouni_2022}, which makes finding an operating point in which the fusion output power is greater than the input power more challenging, since high temperature in a plasma typically means large radiated power in the form of synchrotron and bremsstrahlung radiation. At such high temperatures, bremsstrahlung radiation losses become substantial, and it has been widely argued that fusion power production cannot overcome bremsstrahlung-dominated energy losses under thermonuclear conditions \cite{McNally_1970,Moreau1977,MARTINEZVAL1996,Nevins1998,Nevins2000}. 

Be that as it may, the p-$^{11}$B path to fusion is still attractive since the engineering of the reactor is expected to be much simpler, which in turn trades downstream for current physics challenges. Furthermore, in one way or another, the physics challenges can be overcome sooner or later. Ref.~\cite{Putvinski2019} demonstrated that if properly accounting for kinetic effects, a thermal p-$^{11}$B plasma can produce a high $Q$ (=fusion power/input power), and even reach ignition. By employing a plasma with a low internal magnetic field, and operating in a regime where the
electrons are kept at a lower temperature than the ions, the radiation losses can be further reduced~\cite{Dawson1981}; while maintaining a non-equilibrium population of energetic reacting ions, the fusion power further increases~\cite{Hay_2015}. Using muons to catalyze is another possible option~\cite{Wu2024,Wu:2025bnm}.

These assessments depend critically on the adopted reaction cross-section and the associated thermonuclear reactivity, which directly determine the achievable fusion power density. In recent years, new experimental measurements have significantly extended and refined the available p-$^{11}$B cross-section data, particularly in energy regions that were previously poorly constrained. Notably, measurements by Mazzucconi \emph{et al.}~\cite{Mazzucconi2025} provide high-precision data over a broad energy range and report a previously unobserved resonance structure near 4.7~MeV. This improved experimental landscape highlights the limitations of existing analytic parameterizations, which do not fully capture the updated cross-section behavior and the newly identified resonance features \cite{Nevins2000,Tentori2023}.

We develop a new set of high-accuracy analytic parameterization to the p-$^{11}$B reaction cross-section over the energy range 0--10~MeV in this study, incorporating the latest experimental data. Using these fits, we evaluate the thermonuclear reactivity and quantify the contributions of the dominant resonance at 0.6~MeV and the newly observed resonance near 4.7~MeV. The resulting reactivity is then used to reassess the fusion power density and electron bremsstrahlung power density under self-consistent thermal conditions, enabling a detailed energy balance analysis. Our results show that p-$^{11}$B fusion is not precluded by bremsstrahlung constraints when the energy balance is evaluated using updated cross-section data and self-consistent electron and ion temperatures.

This paper is organized as follows. Sec.~\ref{cs} presents the analytic fitting procedure, data selection, and uncertainty analysis for the p-$^{11}$B cross-section. In Sec.~\ref{Reactivity}, we evaluate the thermonuclear reactivity, focusing on the newly observed resonance near 4.7~MeV and the dominant 0.6~MeV resonance. Sec.~\ref{eb} analyzes the fusion power density and bremsstrahlung losses to assess energy balance conditions. Conclusions are given in Sec.~\ref{conclusion}.


\section{S-factor and cross-section}
\label{cs}
The reaction cross-section $\sigma(E)$, as a function of the center-of-mass energy $E$, is the primary experimental observable and reflects a convolution of nuclear interaction strength, wave-function geometry, and Coulomb barrier penetration.
Following the standard practice adopted in previous evaluations \cite{Nevins2000,Tentori2023}, the experimental cross-section data are transformed into the astrophysical $S$-factor, defined as
\begin{equation}
\sigma(E) = \frac{S(E)}{E} \exp\!\left(-\sqrt{\frac{E_{\mathrm{G}}}{E}}\right),
\label{SE}
\end{equation}
where the exponential term $\exp\left(-\sqrt{E_{\mathrm{G}}/E}\right)$ represents the Gamow factor associated with Coulomb barrier penetration/tunneling, the $1/E$ factor accounts for geometric effects of the wave function, and $E_{\mathrm{G}} = 22.589\,\mathrm{MeV}$ is the Gamow energy of the p-$^{11}$B system \cite{Rolfs1988,Nevins2000}. With this definition, $S(E)$ primarily reflects the nuclear interaction strength and provides a more transparent framework for analyzing reaction mechanisms. Our fitting procedure is then performed in terms of $S(E)$, and the fitted $S$-factor is subsequently converted to the corresponding cross-section through Eq.~(\ref{SE}).

\subsection{Parameterization}
To parameterize the astrophysical $S$-factor over the energy range 0--10\,MeV, we adopt a piecewise functional form building on the parameterization framework introduced in Refs.~\cite{Becker1987,Nevins2000,Tentori2023}. The $S$-factor is expressed as
\begin{equation}
S(E) =
\begin{cases}
S_1(E), & E \leqslant \mathcal{E}_1, \\
S_2(E), & \mathcal{E}_1 < E \leqslant \mathcal{E}_2, \\
S_3(E), & \mathcal{E}_2 < E \leqslant \mathcal{E}_3,
\end{cases}\label{s123}
\end{equation}
with $\mathcal{E}_1=0.4$ MeV, $\mathcal{E}_2$=0.7 MeV and $\mathcal{E}_3$=10.0 MeV being the segment boundaries.

For $E \leqslant \mathcal{E}_1$, the functional form is written as,
\begin{equation}
\begin{aligned}
S_1(E) = {} & C_0 + C_1\left(\frac{E}{1\,\mathrm{keV}}\right)
+ C_2\left(\frac{E}{1\,\mathrm{keV}}\right)^2 \\
& + \frac{A_{\mathrm{L}}}{\left(\frac{E - E_{\mathrm{L}}}{1\,\mathrm{keV}}\right)^2
+ \left(\frac{\delta E_{\mathrm{L}}}{1\,\mathrm{keV}}\right)^2},
\end{aligned}\label{s1}
\end{equation}
where the last Breit-Wigner term represents the narrow resonance at 148~keV, and $C_0$ etc are parameters, similarly hereinafter.

For $\mathcal{E}_1 < E \leqslant \mathcal{E}_2$, the $S$-factor is represented by a polynomial expansion,
\begin{equation}
\begin{aligned}
S_2(E) = {} & D_0 + D_1\left(\frac{E - \mathcal{E}_1}{100\,\mathrm{keV}}\right)
+ D_2\left(\frac{E - \mathcal{E}_1}{100\,\mathrm{keV}}\right)^2 \\
& + D_5\left(\frac{E - \mathcal{E}_1}{100\,\mathrm{keV}}\right)^5.
\end{aligned}\label{s2}
\end{equation}

For $\mathcal{E}_2 < E \leqslant \mathcal{E}_3$, the fitting function is constructed as a sum of Breit-Wigner resonance terms,
\begin{equation}
S_3(E) = B + \sum_{k=0}^{4}
\frac{A_k}{\left(\frac{E - E_k}{1\,\mathrm{keV}}\right)^2
+ \left(\frac{\delta E_k}{1\,\mathrm{keV}}\right)^2},\label{s3}
\end{equation}
here, we have introduced an additional resonance term ($k=4$) to account for the new resonance structure observed near 4.7~MeV in recent measurements~\cite{Mazzucconi2025}.

\subsection{Data Selection}
To evaluate the p-$^{11}$B reaction cross-section over the energy interval 0--10\,MeV, we construct the parameterization using three experimental datasets: Becker \emph{et al.}~\cite{Becker1987}, Mazzucconi \emph{et al.}~\cite{Mazzucconi2025}, and Buck \emph{et al.}~\cite{Buck1983}.

We adopt the dataset of Becker \emph{et al.}~\cite{Becker1987}, covering 0.022--1.1\,MeV. Their study combined measurements using proton beams incident on solid $^{11}\mathrm{B}$ targets with inverse-kinematics experiments employing $^{11}\mathrm{B}$ beams on gaseous hydrogen targets, providing a standard cross-section reference in this region~\cite{Nevins2000,Tentori2023,Mazzucconi2025,Munch_2020}.

We further adopt the recent measurements of Mazzucconi \emph{et al.}~\cite{Mazzucconi2025}, spanning 0.34--4.73\,MeV, rather than those of the earlier widely used evaluation of Sikora \emph{et al.}~\cite{Sikora2016}. The measurements of Mazzucconi \emph{et al.} supplement existing data in previously explored regions, fill a gap in the 3.85--4.73\,MeV interval, and reveal a resonance structure near 4.7\,MeV. In contrast, Sikora \emph{et al.}~\cite{Sikora2016} analyzed $\alpha$-particle spectra measured by Spraker \emph{et al.}~\cite{Spraker_2012} by normalizing the observed data yields to a theoretical sequential-decay model~\cite{Stave_2011_11B}. As a result, the extracted absolute cross-section is inherently model dependent; for example, different assumptions for the orbital angular momentum of the primary $\alpha$ emission (e.g.\ $\ell=1$ versus $\ell=3$) lead to substantially different normalization factors. By contrast, Mazzucconi \emph{et al.} employed a monolithic silicon telescope that enables event-by-event discrimination between reaction-produced $\alpha$ particles and scattered protons. This technique allows the complete $\alpha$ spectrum to be measured directly, without model-based subtraction or extrapolation. Mazzucconi \emph{et al.} provide a more direct determination of the reaction cross-section in this energy interval, reducing a major source of systematic uncertainty.

To extend the evaluation to the energy range from 5 to 10\,MeV, we incorporate the measurements of Buck \emph{et al.}~\cite{Buck1983}. Their results show that, over this interval, the reaction cross-section decreases rapidly with increasing energy and becomes small compared with its values at lower energies. The overall energy dependence in this range is therefore the primary constraint provided by these data, rather than the precise absolute normalization~\cite{Tentori2023,Mazzucconi2025}.

\subsection{Results}
\begin{figure*}
	\centering
	\includegraphics[width=0.8\linewidth]{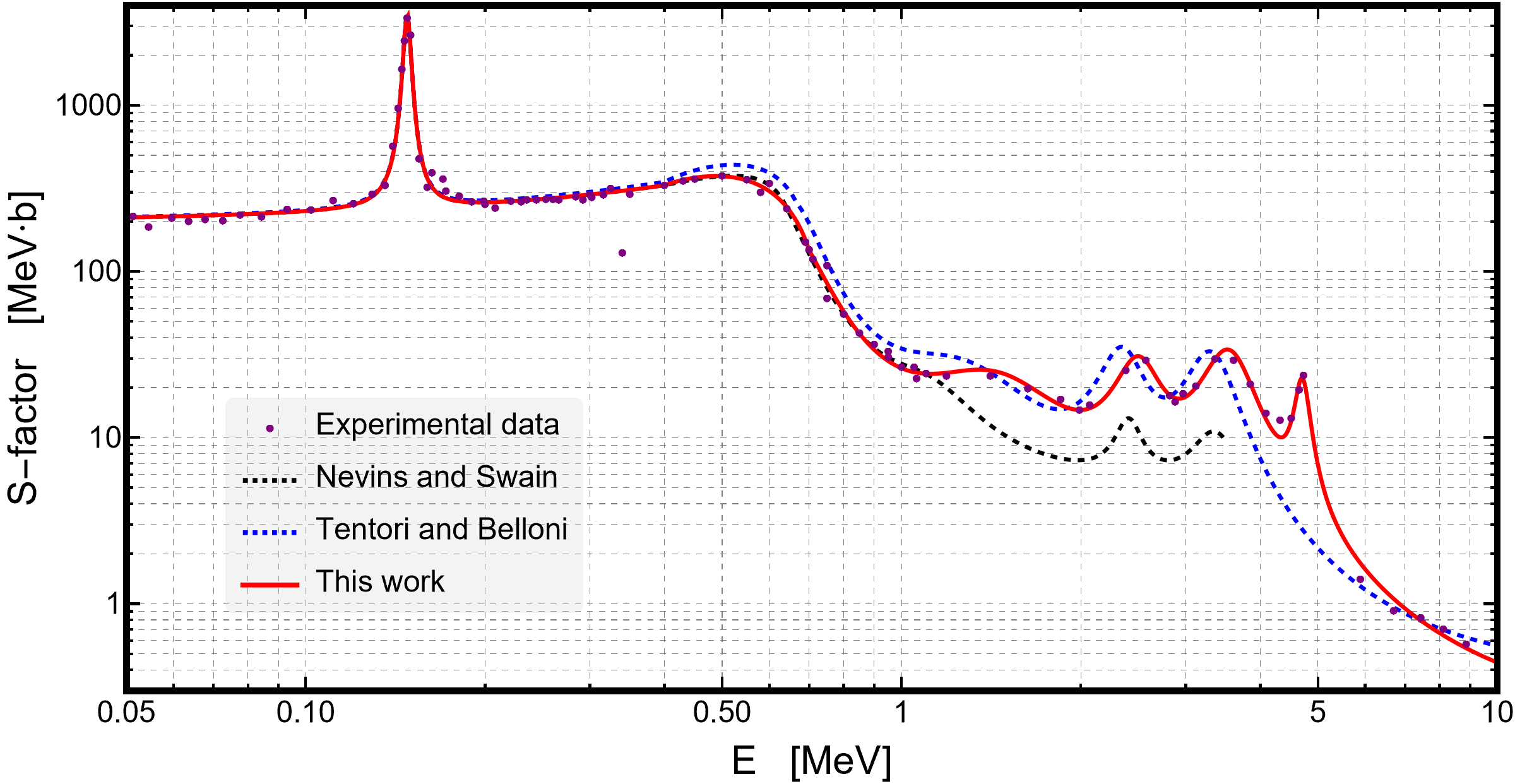}
    \caption{
Astrophysical \(S\)-factor of the p-$^{11}$B fusion reaction as a function of the center-of-mass energy.
Experimental data are taken from Becker \emph{et al.}~\cite{Becker1987} and Mazzucconi \emph{et al.}~\cite{Mazzucconi2025} for \(E \leq 5\)~MeV, and from Buck \emph{et al.}~\cite{Buck1983} for \(E \geq 5\)~MeV.
The analytic parameterization frame used herein, which is defined by Eqs.~(\ref{s123})-(\ref{s3}), is shown together with previous parameterizations by Nevins and Swain~\cite{Nevins2000} and Tentori \emph{et al.}~\cite{Tentori2023} for comparison.
}
    \label{fig:Sfactors}
\end{figure*}

The fitted $S(E)$ is shown in Fig.~\ref{fig:Sfactors}, with parameters obtained from weighted least-squares minimization and summarized in Table~\ref{tab:fit_params}.

\begin{table}
\centering
\caption{Fitting parameters for the p-$^{11}$B reaction $S$-factor}
\label{tab:fit_params}
\begin{minipage}[t]{0.48\linewidth}
\centering
\begin{tabular}{lc}
\toprule
Parameter & Value \\
\midrule
\multicolumn{2}{l}{$E \leq 0.4$ MeV} \\
\midrule
$C_0$ (MeV b) & 197 \\
$C_1$ (MeV b) & 0.240 \\
$C_2$ (MeV b) & $2.31\times10^{-4}$ \\
$A_L$ (MeV b) & $1.82\times10^{4}$ \\
$E_L$ (keV) & 148 \\
$\delta E_L$ (keV) & 2.35 \\
\midrule
\multicolumn{2}{l}{$0.4 < E \leq 0.7$ MeV} \\
\midrule
$D_0$ (MeV b) & 330.2 \\
$D_1$ (MeV b) & 102.436 \\
$D_2$ (MeV b) & -58.481 \\
$D_5$ (MeV b) & 0.0933 \\
\bottomrule
\end{tabular}
\end{minipage}
\hfill
\begin{minipage}[t]{0.48\linewidth}
\centering
\begin{tabular}{lc}
\toprule
Parameter & Value \\
\midrule
\multicolumn{2}{l}{$0.7 < E \leq 10.0$ MeV} \\
\midrule
$B$ (MeV b) & 0.209689 \\
$A_0$ ($\times10^6$ MeV b) & 2.0235 \\
$A_1$ ($\times10^6$ MeV b) & 4.0102 \\
$A_2$ ($\times10^6$ MeV b) & 1.3220 \\
$A_3$ ($\times10^6$ MeV b) & 4.9451 \\
$A_4$ ($\times10^5$ MeV b) & 4.3430 \\
$E_0$ (MeV) & 0.6222 \\
$E_1$ (MeV) & 1.3884 \\
$E_2$ (MeV) & 2.4924 \\
$E_3$ (MeV) & 3.5286 \\
$E_4$ (MeV) & 4.7036 \\
$\delta E_0$ (MeV) & 0.0996 \\
$\delta E_1$ (MeV) & 0.4499 \\
$\delta E_2$ (MeV) & 0.2386 \\
$\delta E_3$ (MeV) & 0.3985 \\
$\delta E_4$ (MeV) & 0.1525 \\
\bottomrule
\end{tabular}
\end{minipage}
\end{table}

In the energy interval below 0.4\,MeV (the isolated point at $E$=0.34 MeV is excluded), the fitted curve follows the Nevins-Swain parameterization~\cite{Nevins2000}, as new datasets introduced no experimental discrepancies in this range. The resulting $S$-factor remains consistent with existing experimental data.

Over the interval 0.4--0.7\,MeV, the fit reproduces the experimental $S(E)$ with high fidelity. The coefficient of determination reaches $R^2=0.97$, and the mean relative deviation is approximately 3\%, well within the experimental uncertainty reported by Mazzucconi \emph{et al.}~\cite{Mazzucconi2025}. Larger local deviations occur near 0.6\,MeV, reflecting increased experimental dispersion in the vicinity of the dominant resonance, consistent with previous analyses~\cite{Tentori2023}.

Across the range 0.7--5.0\,MeV, the fit quality remains comparable, with $R^2=0.97$ and a mean relative deviation of 6.2\%, under the experimental uncertainty. In this sense, the inclusion of an additional resonance term we introduced in Eq.~(\ref{s3}), successfully accounts for the structure observed near 4.7\,MeV and improves the description of the experimental data in this interval.

For energies between 5 and 10\,MeV, the fitted $S$-factor reproduces the overall trend of the data, although the coefficient of determination decreases to $R^2=0.62$. This reflects the sparse and discrete nature of the available measurements~\cite{Buck1983}. The mean relative deviation of 12.4\% represents a substantial reduction compared with earlier evaluations~\cite{Tentori2023}, attributable to the extended resonance parameterization adopted here.

\begin{figure*}
	\centering
	\includegraphics[width=0.8\linewidth]{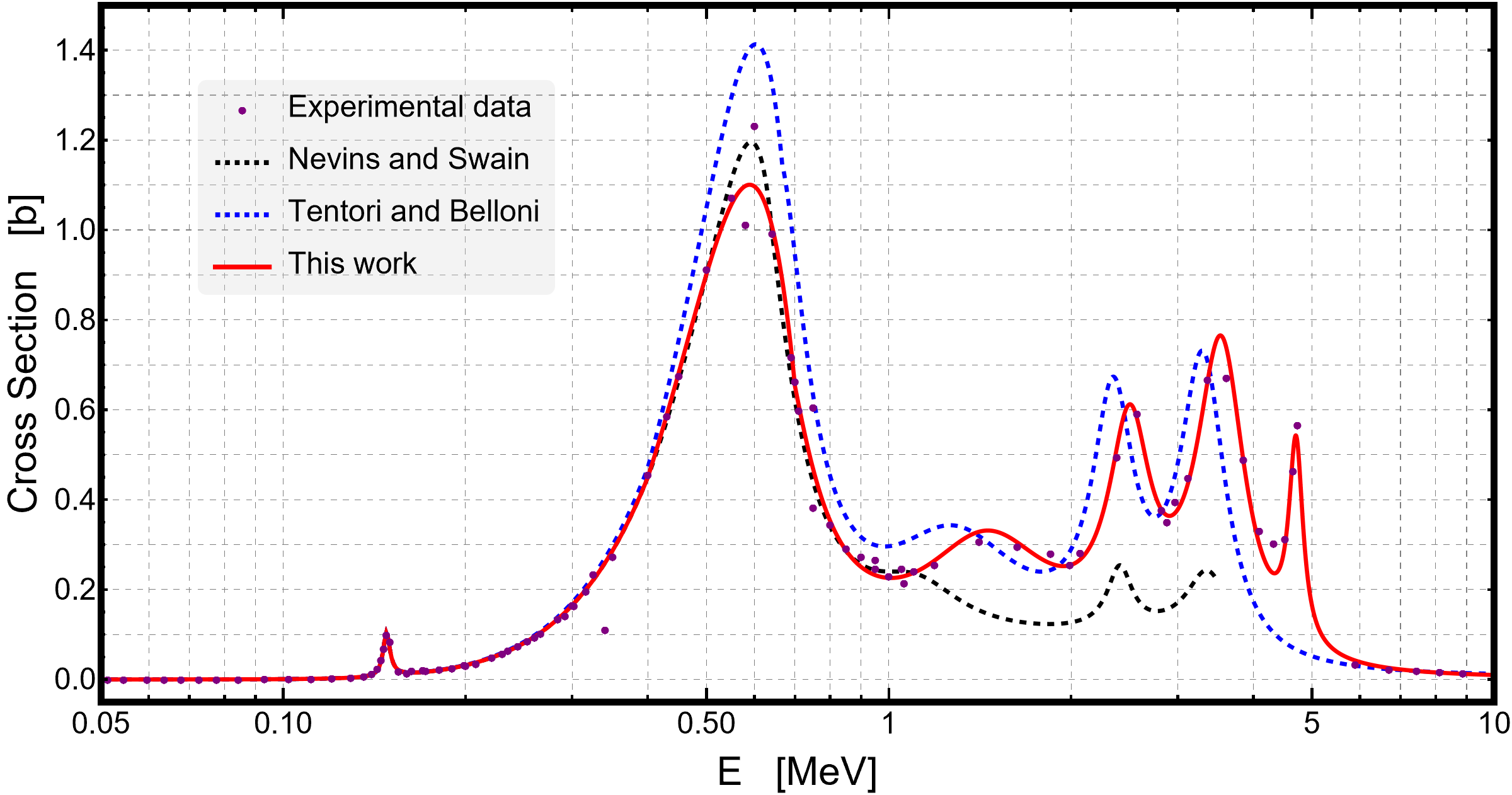}
    \caption{
p-$^{11}$B fusion reaction cross-section $\sigma(E)$ as a function of the center-of-mass energy, obtained by transforming the fitted astrophysical \(S\)-factor using Eq.~(\ref{SE}).
The cross-section corresponds to the same parameter set shown in Fig.~\ref{fig:Sfactors}.
} \label{fig:cross_sections}
\end{figure*}

The corresponding reaction cross-sections, obtained by transforming the fitted $S(E)$ via Eqs.~(1), are shown in Fig.~\ref{fig:cross_sections}. The same parameter set is used, and the quality of the approximation follows directly from that of the $S$-factor fit.

Overall, the present parameterization provides an improved representation of recent experimental data while preserving the physical structure of the Nevins-Swain framework, forming a consistent basis for the subsequent reactivity calculations.


\section{Reactivity}
\label{Reactivity}
Both ignition conditions and achievable fusion power density at fixed plasma pressure are governed by the thermonuclear reactivity. An accurate evaluation of the p-$^{11}$B reactivity over a broad ion-temperature range is therefore essential for assessing its feasibility as a fusion fuel.

The thermonuclear reactivity $\langle\sigma v\rangle$ is obtained by convolving the reaction cross-section with a Maxwellian velocity distribution and is given by~\cite{Rolfs1988}
\begin{equation}
\langle\sigma v\rangle = \frac{(8/\pi)^{1/2}}{\mu^{1/2}(k_B T)^{3/2}}
\int_{0}^{\infty} \sigma E \exp\!\left(-E/k_B T\right)\, dE ,
\end{equation}
where $\mu$ is the reduced mass of the p-$^{11}$B system, and $T$ is the ion temperature.

Using the cross-section parameterization developed in Sec.~\ref{cs}, we evaluate the temperature dependent reactivity and quantify the respective impacts of the newly observed resonance near 4.7\,MeV and the dominant 0.6\,MeV resonance.

\subsection{4.7 MeV Resonance}
The resonance structure observed near 4.7\,MeV represents a qualitatively new feature of the p-$^{11}$B reaction cross-section. While its contribution to the reactivity at lower ion temperatures is negligible, its influence at elevated temperatures has not been systematically quantified.

To isolate the effect of this resonance, two cases are considered: (i) the full parameterization including all resonance terms ($k=0$--4), and (ii) an otherwise identical parameterization with the 4.7\,MeV resonance removed ($k=0$--3). This comparison allows a direct assessment of the resonance-induced modification of the Maxwellian-averaged reactivity.

\begin{figure}
	\centering
	\includegraphics[width=\linewidth]{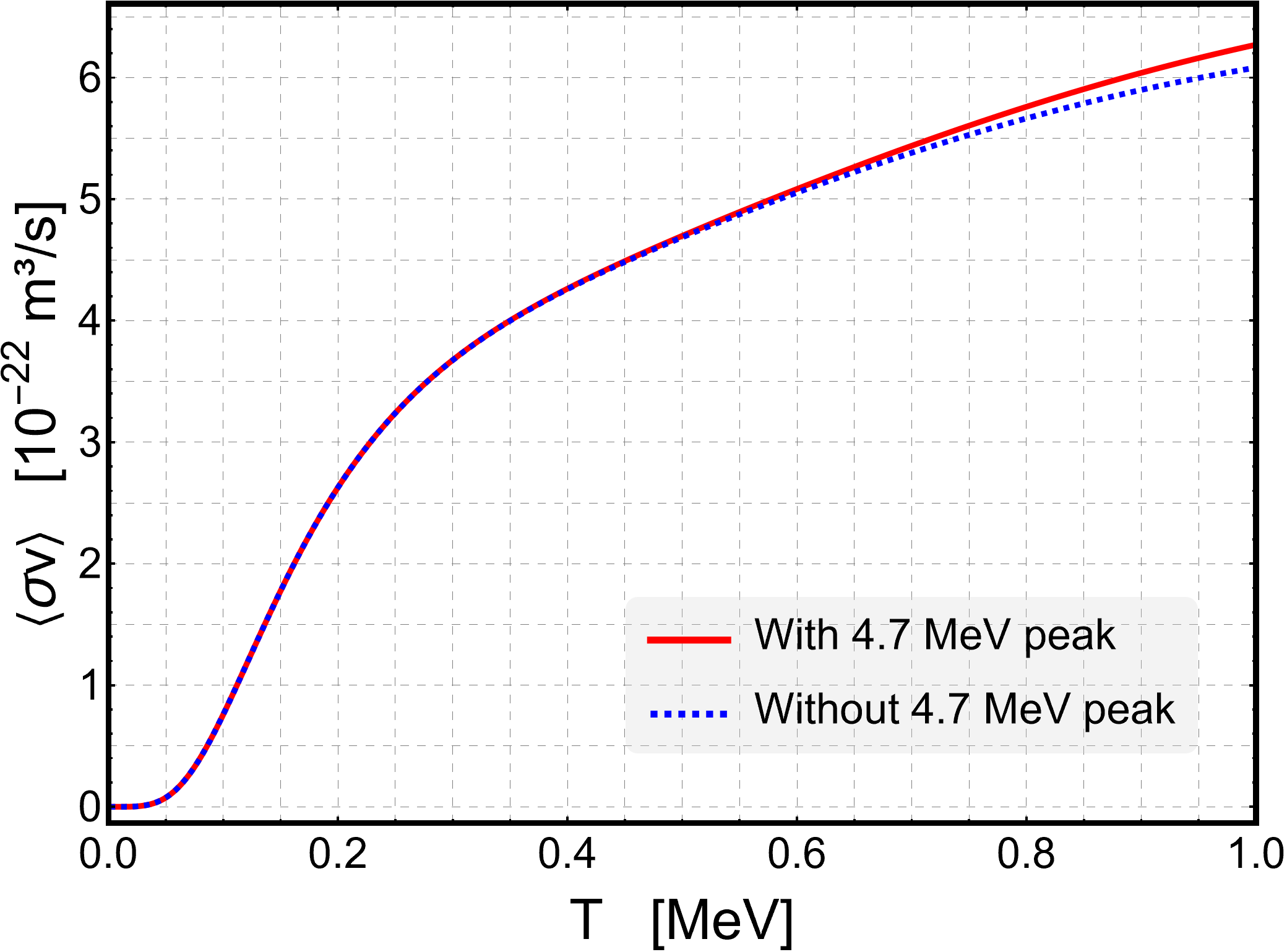}
    \caption{
Thermonuclear reactivity of the p-$^{11}$B reaction as a function of ion temperature.
The red solid curve shows the result obtained with inclusion of the resonance structure near 4.7~MeV, while the blue dashed curve corresponds to the calculation with this resonance removed.
}  \label{fig:reactivity_4.7}
\end{figure}

\begin{figure}
	\centering
	\includegraphics[width=\linewidth]{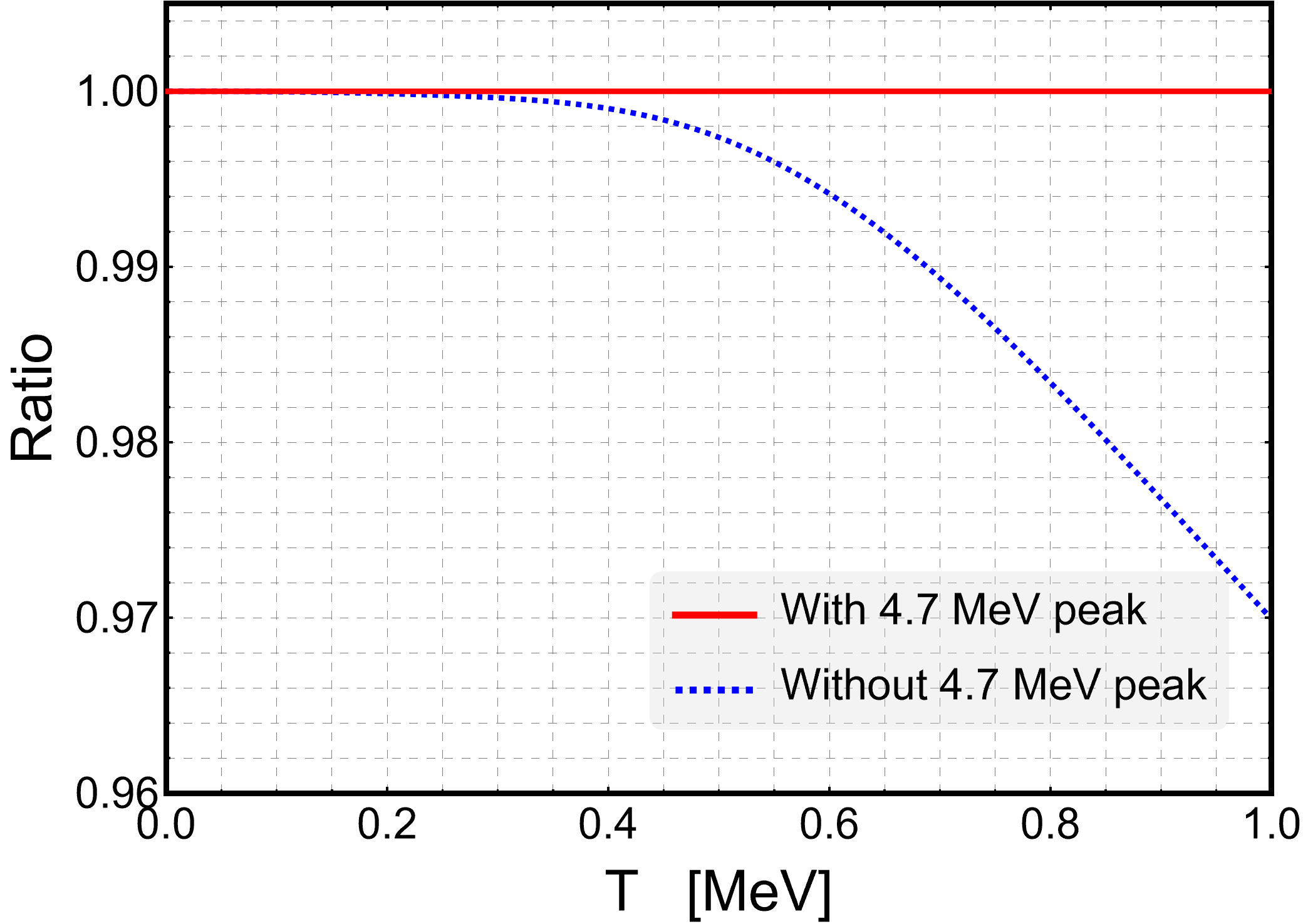}
    \caption{
Ratio of the thermonuclear reactivity calculated without the 4.7~MeV resonance to that obtained with the full cross-section parameterization, shown as a function of ion temperature.
} \label{fig:ratio4.7}
\end{figure}

As shown in Fig.~\ref{fig:reactivity_4.7}, the impact of the 4.7\,MeV resonance is negligible for $T < 0.4$\,MeV, reflecting the exponential suppression of contributions from higher center-of-mass energies at lower temperatures. This behavior is consistent with earlier observations that p-$^{11}$B reactivity exhibits weak sensitivity to high-energy cross-section features in this temperature range~\cite{Sikora2016}. However, the impact of the 4.7\,MeV resonance becomes increasingly pronounced at ion temperatures above $0.4$\,MeV. In this regime, the increased contribution from the upper-energy part of the cross-section in the reactivity integral leads to a measurable enhancement of $\langle\sigma v\rangle$, as more clearly shown in Fig.~\ref{fig:ratio4.7}.

\subsection{0.6 MeV Resonance}
The resonance near 0.6\,MeV constitutes the dominant structural feature of the p-$^{11}$B cross-section in the energy range most relevant for thermonuclear applications. Consequently, uncertainties in its magnitude directly propagate into uncertainties in the calculated reactivity. Current experimental measurements in this region exhibit uncertainties at the $\sim10\%$ level~\cite{Mazzucconi2025,Munch_2020,Spraker_2012,Sikora2016}.

\begin{figure}
	\centering
	\includegraphics[width=\linewidth]{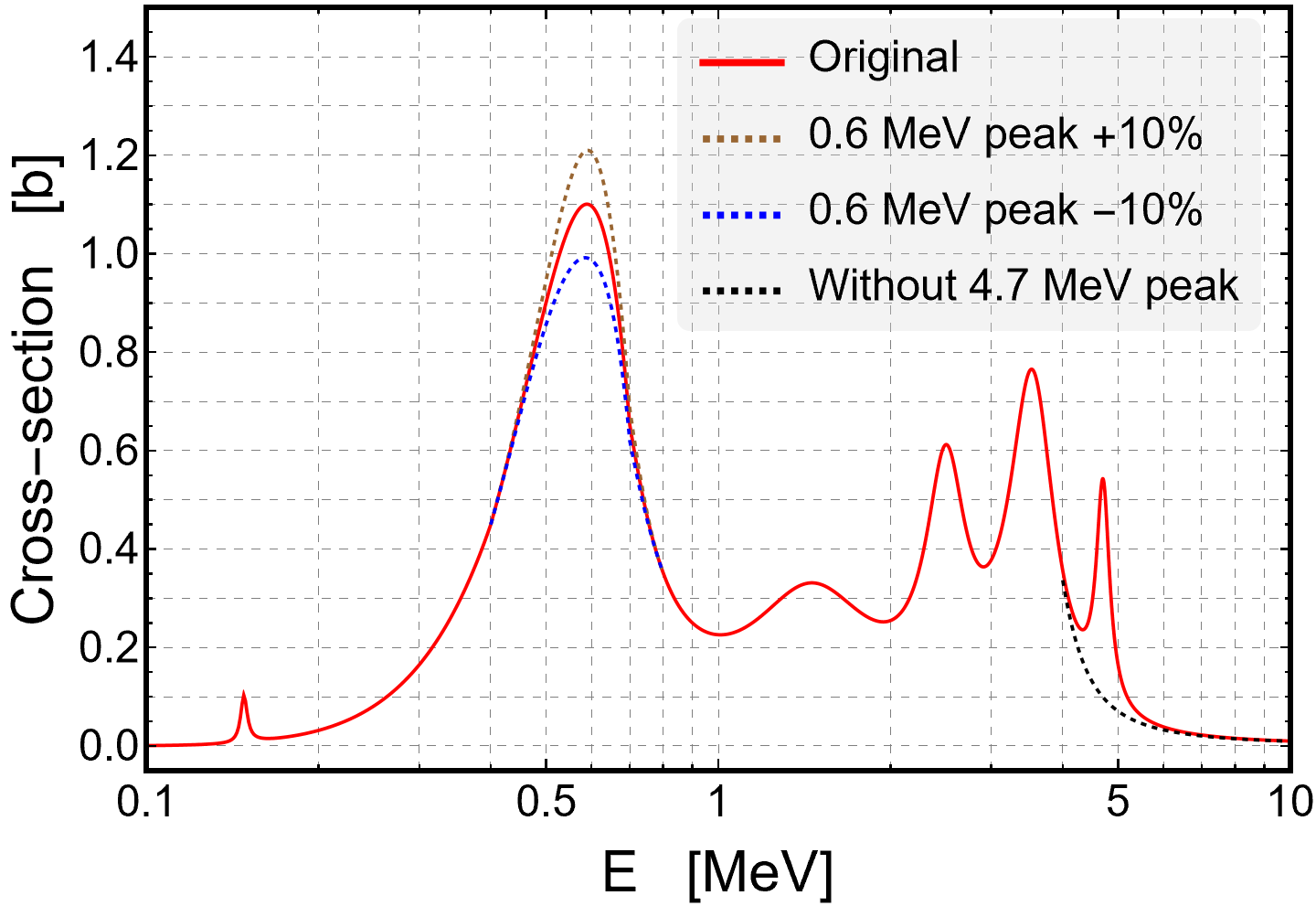}
    \caption{
Illustrative variations of the p-$^{11}$B reaction cross-section associated with the dominant resonance near 0.6~MeV and the newly observed resonance around 4.7~MeV.
These variations are used to assess the sensitivity of the thermonuclear reactivity to uncertainties in the resonance structures.
} \label{fig:cross_section_adjustments}
\end{figure}

To quantify the resulting sensitivity, we perform a systematic variation of the 0.6\,MeV resonance strength, as shown in Fig.~\ref{fig:cross_section_adjustments}. Modified cross-section distributions are constructed by smoothly increasing and decreasing the peak cross-section by $\pm10\%$ over the 0.4--0.8\,MeV interval. Here we introduce a cubic transition function,
\begin{equation}
f(t) = 3t^2 - 2t^3, \qquad
t = \frac{x - x_{\text{start}}}{x_{\text{end}} - x_{\text{start}}},
\end{equation}
to ensure continuity of both the cross-section and its first derivative, with $x_{\text{start}}=0.4,\,0.6$\,MeV and $x_{\text{end}}=0.6,\,0.8$\,MeV, respectively.

\begin{figure}
	\centering
	\includegraphics[width=\linewidth]{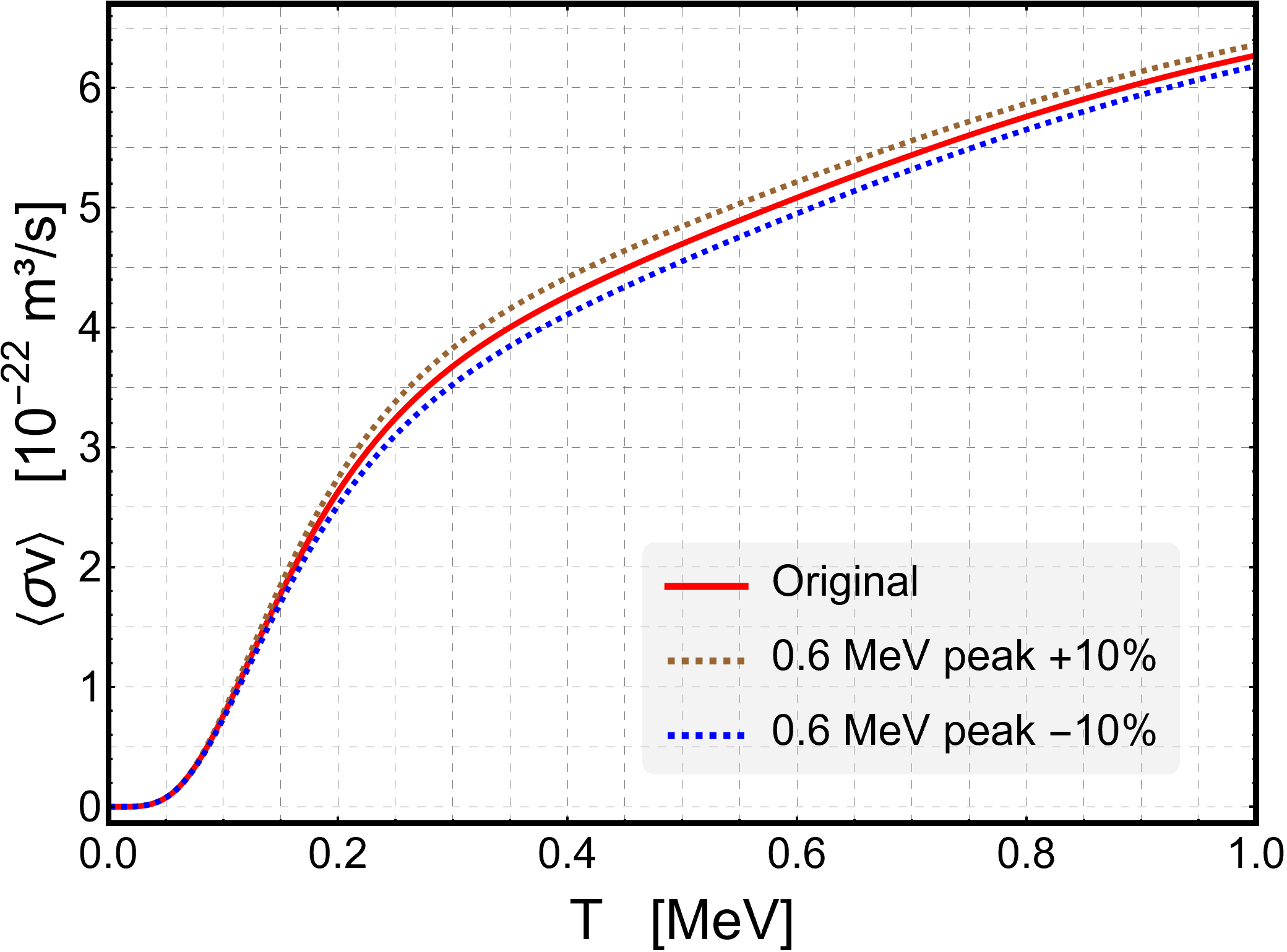}
    \caption{
Thermonuclear reactivity of the p-$^{11}$B reaction as a function of ion temperature, obtained after upward and downward adjustments of the peak strength of the dominant 0.6~MeV resonance by \(\pm10\%\).
} \label{fig:reactivity_0.6}
\end{figure}

\begin{figure}
	\centering
	\includegraphics[width=\linewidth]{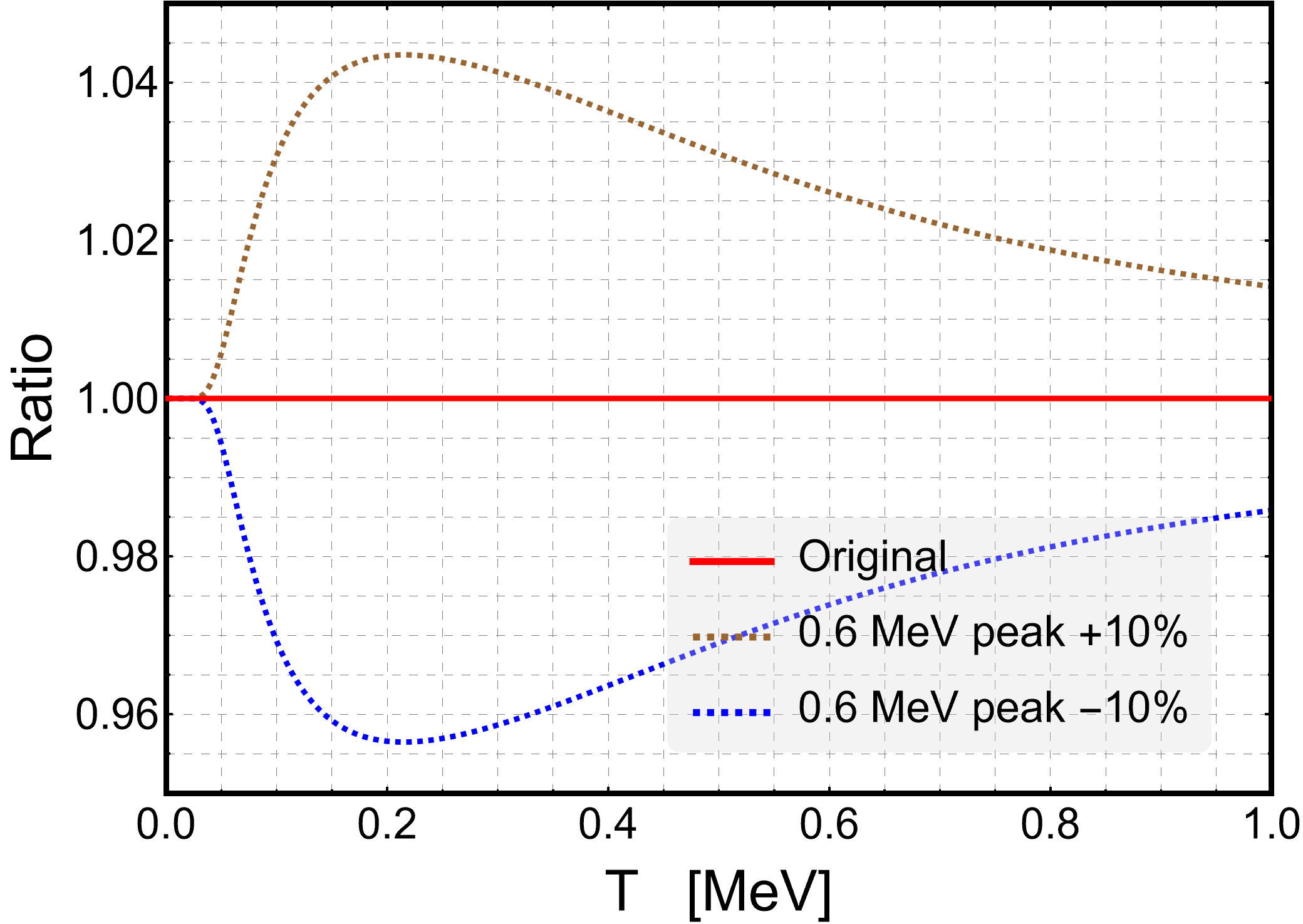}
    \caption{
Ratio of the thermonuclear reactivity corresponding to \(\pm10\%\) variations in the strength of the 0.6~MeV resonance to the reactivity calculated using the original, unmodified cross-section parameterization, shown as a function of ion temperature.
} \label{fig:ratio0.6}
\end{figure}

As we can see in Fig.~\ref{fig:reactivity_0.6}, variations in the 0.6\,MeV resonance strength have a pronounced impact on the reactivity over a broad temperature range. Within the interval $T=0.1$--0.4\,MeV, a $\pm10\%$ change in the peak cross-section results in a 3--5\% variation in $\langle\sigma v\rangle$. In particular, near $0.2$\,MeV, the difference between the $+10\%$ and $-10\%$ cases produces a total spread in reactivity approaching 10\%, as drawn in Fig.~\ref{fig:ratio0.6}. This level of variation is comparable to that reported in existing reactivity evaluations based on different experimental datasets~\cite{Nevins2000,Putvinski2019,Tentori2023}. This observation indicates that uncertainties associated with the 0.6\,MeV resonance constitute a major contributor, among others, to the spread among current p-$^{11}$B reactivity models.

\section{Energy Balance}
\label{eb}
In thermonuclear fusion systems, ignition and sustained burn require that fusion-generated heating compensate for the energy-loss mechanisms. For the p-$^{11}$B reaction, this condition is particularly restrictive owing to the strong electron bremsstrahlung associated with high-$Z$ ions. An assessment of the energy balance is therefore essential for evaluating the viability of p-$^{11}$B fusion.

In this section, we examine the energy balance by comparing the fusion power density with the electron bremsstrahlung power density, using the reactivity results obtained in Sec.~\ref{Reactivity}.

\subsection{Model}
In the p-$^{11}$B reaction, nearly all of the released energy is carried by energetic $\alpha$ particles. Through Coulomb interactions, this energy is predominantly transferred to the ion population, with only a small fraction deposited directly into electrons. Following the conventional treatment adopted in previous studies~\cite{Dawson1981,Nevins1998,Putvinski2019}, we also assume that the fusion energy is entirely transferred to the ions.

Under this assumption, the fusion power density is
\begin{equation}
P_{\mathrm{fusion}} = n_e^2 f_1 f_2 \langle \sigma v \rangle Y ,
\end{equation}
where $n_e$ is the electron density, $f_1 = n_1/n_e$ and $f_2 = n_2/n_e$ denote the relative densities of protons and $^{11}$B ions, respectively, and $Y = 8.7~\mathrm{MeV}$ is the energy released per fusion reaction.

Since both fusion power and bremsstrahlung losses scale as $n_e^2$, we consider the normalized fusion power density
\begin{equation}
\frac{P_{\mathrm{fusion}}}{n_e^2} = f_1 f_2 \langle \sigma v \rangle Y ,
\end{equation}
which isolates the intrinsic temperature dependence of the reaction. Throughout this analysis, we adopt the reference composition used by Nevins~\cite{Nevins1998}, with $f_1 = 1/2$ and $f_2 = 1/10$.

Among the various energy-loss channels, including synchrotron radiation, transport losses, and thermal conduction, electron bremsstrahlung is known to dominate under the conditions relevant to p-$^{11}$B fusion~\cite{Fowler1967,Fowler1975,Dawson1981,Nevins1998}. We therefore retain bremsstrahlung as the primary loss mechanism.

The bremsstrahlung power density is evaluated using the analytic expression proposed by Rider~\cite{Rider1995,Rider1997}, and is written in normalized form as
\begin{equation}
\begin{aligned}
\frac{P_{\mathrm{brem}}}{n_e^2}
= {} & 5.172 \times 10^{-43} \sqrt{T_e}
\Bigg\{
Z_{\mathrm{eff}}
\left[
1 + 0.7936 \frac{T_e}{m_e c^2}
\right. \\
& \left.
+ 1.874 \left( \frac{T_e}{m_e c^2} \right)^2
\right]
+ \frac{3}{\sqrt{2}} \frac{T_e}{m_e c^2}
\Bigg\} .
\end{aligned}\label{brem_pd}
\end{equation}
in units of $\mathrm{MW\,m^3}$, where $T_e$ is the electron temperature and
$Z_{\mathrm{eff}} = \sum_i n_i Z_i^2 / \sum_i n_i Z_i$ is the effective charge number.

The electron temperature is determined self-consistently from the electron energy balance. Electron heating is dominated by energy transfer from thermal ions, while electron energy losses are primarily due to bremsstrahlung. The electron-ion energy transfer rate is given by
\begin{equation}
P_{\mathrm{trans}}(T, T_e)
= \frac{3}{2} \nu_{\mathrm{trans}} (T - T_e),
\end{equation}
where $\nu_{\mathrm{trans}}$ is the electron-ion energy exchange coefficient (see Eq. (2.17) in~\cite{Braginski_1961}). The electron temperature $T_e$ is obtained by solving the energy balance equation:
\begin{equation}
P_{\mathrm{trans}}(T, T_e) = P_{\mathrm{brem}}(T_e).
\end{equation}
By substituting  the calculated electron temperature into Eq.~(\ref{brem_pd}), we obtain the normalized bremsstrahlung power density as a function of ion temperature.

\subsection{Results}
\begin{figure}
	\centering
	\includegraphics[width=\linewidth]{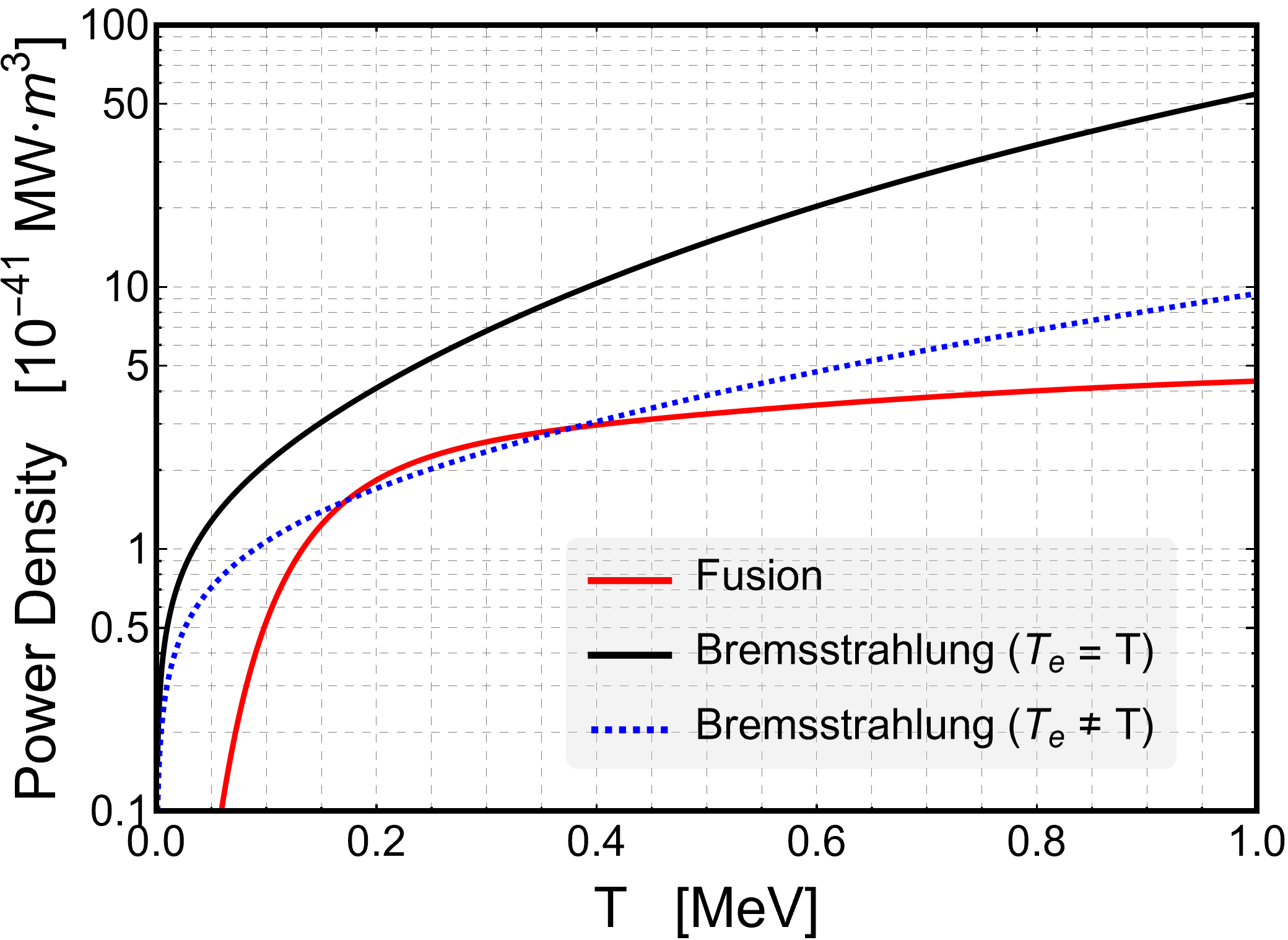}
    \caption{
Normalized fusion power density and electron bremsstrahlung power density as functions of ion temperature.
Results are shown for the assumption of electron-ion thermal equilibrium (\(T_e = T\)) and for a self-consistent electron temperature determined from the electron energy balance equation.
} \label{fig:fusion_brem}
\end{figure}

Fig.~\ref{fig:fusion_brem} compares the normalized fusion power density and bremsstrahlung power density as functions of ion temperature. The fusion power density is shown by the red curve. Obviously, if electron-ion thermal equilibrium is assumed ($T_e = T$), the resulting bremsstrahlung loss (black curve) exceeds the fusion power density over the entire temperature range considered, consistent with earlier assessments~\cite{Nevins1998}. However, this assumption is not generally valid for p-$^{11}$B plasmas, where electrons and ions may be significantly out of thermal equilibrium.

When the electron temperature is determined self-consistently from the energy balance equation, the resulting bremsstrahlung power density (blue dashed curve) is substantially reduced. In this case, two intersections between fusion power and bremsstrahlung loss appear at $T \simeq 0.17$ and $0.38~\mathrm{MeV}$, indicating a temperature window in which fusion power exceeds bremsstrahlung losses. This behavior is consistent with recent analyses~\cite{Putvinski2019}, and suggests that p-$^{11}$B fusion is not fundamentally excluded by bremsstrahlung constraints when self-consistent electron temperatures are employed.

\begin{figure}
	\centering
	\includegraphics[width=\linewidth]{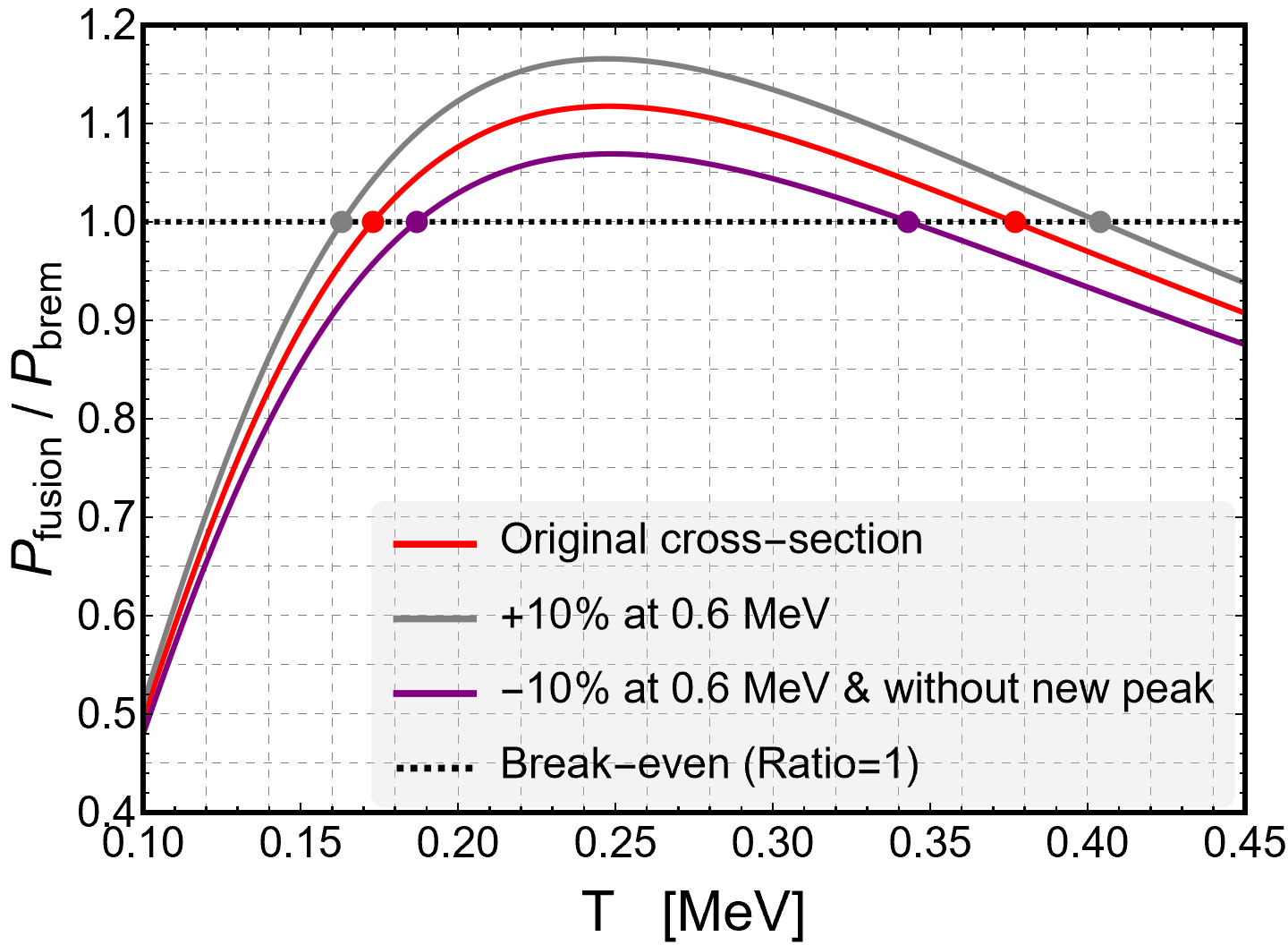}
    \caption{
Ratio of normalized fusion power density to electron bremsstrahlung power density as a function of ion temperature.
The result obtained with the original cross-section parameterization is compared with bounding cases corresponding to \(\pm10\%\) variations in the strength of the 0.6~MeV resonance, with the conservative case also excluding the 4.7~MeV resonance.
} \label{fig:fusion_brem_ratio}
\end{figure}

To assess the sensitivity of this conclusion to uncertainties in the reaction cross-section, we also consider two bounding cases, as depicted in Fig.~\ref{fig:fusion_brem_ratio}. An optimistic case (gray curve) is constructed by increasing the 0.6~MeV resonance strength by $10\%$, while a conservative case (purple curve) is obtained by reducing the same resonance by $10\%$ and omitting the newly observed 4.7~MeV resonance, shown in Fig.~\ref{fig:cross_section_adjustments}. In both cases, the qualitative structure of the energy-balance window is preserved, with fusion power exceeding bremsstrahlung losses over a comparable temperature interval and peaking near $T \simeq 0.25~\mathrm{MeV}$.

These results indicate that, within current experimental uncertainties, the p-$^{11}$B reaction admits a temperature range in which a positive net power balance is achievable, consistent with ignition estimates reported in previous theoretical studies~\cite{Watanabe_2004,Hay_2015,Ochs_2022}.
Furthermore, as discussed in the Introduction, a low internal magnetic field, keeping the
electrons at a lower temperature than the ions, maintaining a non-equilibrium population of energetic reacting ions, using muons to catalyze, and so on, may all help to make the window larger.


\section{Conclusion}
\label{conclusion}
At present, the majority of energy utilized globally is derived from fossil fuels. Regrettably, these sources are not only detrimental to the environment but also non-renewable. Hence, despite persistent scientific and engineering challenges, the pursuit of controlled nuclear fusion energy has remained a longstanding aspiration of humanity for more than 70 years.

The fusion of a proton with a boron nucleus (p-$^{11}$B) has always been considered the holy grail due to its exceptional potential for commercialization.
Of course, the main rationale is not rooted in simplicity, but rather in the constraints imposed by limited alternatives. The scarcity of available fuels renders options such as deuterium-tritium (D-T) and D-helium 3 ($^3$He) impractical. Furthermore, the necessity to circumvent neutron-related complications also makes D-D fusion an unfavorable option. Per contra, the p-$^{11}$B reaction offers numerous benefits. Notably, the plentiful availability of boron fuel on Earth is anticipated to considerably reduce expenses. Furthermore, the absence of a breeding blanket requirement, which is unavoidable in the D-T reaction, streamlines the design and diminishes the fusion device's overall size. Additionally, its aneutronic characteristics reduce the necessity for a dense screen blanket, thereby further facilitating size reduction. This opens up the possibility of direct energy conversion or direct electric power generation, contingent upon the controlled production of only charged $\alpha$ particles.

In this study, we developed a high-precision analytical parameterization of the p-$^{11}$B reaction cross-section over the center-of-mass energy range 0--10~MeV, incorporating the latest experimental data. Based on this parameterization, we have systematically evaluated the thermonuclear reactivity of the p-$^{11}$B reaction, and quantified the roles of the dominant resonance near 0.6~MeV as well as the newly observed resonance structure around 4.7~MeV. We find that, while the 4.7~MeV resonance represents an important new feature of the p-$^{11}$B reaction cross-section and provides valuable insight into the underlying reaction mechanism, its contribution to the reactivity remains negligible at ion temperatures below $T \simeq 0.4$~MeV. Consequently, since the ion temperature window relevant for positive energy balance lies well below the energies at which the 4.7~MeV resonance contributes appreciably to the reactivity, its inclusion has only a minor influence on the energy balance assessment within the parameter variations considered.

By contrast, the resonance near 0.6~MeV is found to exert a dominant influence on the reactivity over the temperature range of primary interest for thermonuclear applications. A $\pm10\%$ variation in its peak strength leads to systematic changes in the calculated reactivity that are comparable in magnitude to the spread among existing reactivity evaluations based on different experimental datasets. This sensitivity highlights the central role of the 0.6~MeV resonance in determining the quantitative accuracy of present p-$^{11}$B reactivity models.

Using the updated reactivity together with a self-consistent treatment of the electron temperature, we have reassessed the energy balance between fusion power density and electron bremsstrahlung losses. While the assumption of electron-ion thermal equilibrium reproduces the conventional conclusion that bremsstrahlung losses dominate, allowing the electron temperature to be determined by the electron energy balance yields a finite ion-temperature window in which fusion power exceeds bremsstrahlung radiation. This result remains robust under conservative variations of the underlying cross-section data within current experimental uncertainties.

Taken together, these results indicate that p-$^{11}$B fusion, while requiring stringent plasma conditions, is not fundamentally excluded by bremsstrahlung constraints when contemporary cross-section measurements and a self-consistent thermal treatment are employed. Continued progress in experimental nuclear physics, together with improved plasma modeling, will be essential for further reducing uncertainties and refining assessments of p-$^{11}$B fusion as a viable aneutronic energy source.


\begin{acknowledgments}
This work was supported by Natural Science Foundation of Jiangsu Province (grant no. BK20220122); China Postdoctoral Science Foundation (grant no. 2024M751369); National Natural Science Foundation of China (grant no. 12233002), and Jiangsu Funding Program for Excellent Postdoctoral Talent. Use of the computing resources at \href{www.syli.cloud}{Jiangsu Xilixi Technology} is gratefully acknowledged.
\end{acknowledgments}


\nocite{*}
\bibliographystyle{apsrev4-1}
\bibliography{refs} 
\end{document}